\documentclass[a4paper,12pt,oneside,onecolumn]{article}
\usepackage{authblk}
\usepackage {amsfonts}
\usepackage {graphicx}
\usepackage {epsfig}
\usepackage{amssymb}
\usepackage{amsmath}
\usepackage {longtable}
\usepackage[english]{babel}
\graphicspath{{img/}} 
\usepackage{color,soul}
\usepackage{todonotes}
\usepackage{cite}

	\title{About possibility to observe spin dichroism effect (the effect of tensor polarization acquiring) for
		a nonpolarized deuteron beam passing through the nonpolarized internal target of Nuclotron		}
	
\author[1]{S.V.~Anischenko}
\author[1]{V.G.~Baryshevsky}
\author[1]{A.A.~Gurinovich}
	
\author[2]{V. P. Ladygin }

\affil[1]{Institute for Nuclear Problems, Minsk, Belarus}
\affil[2]{Joint Institute for Nuclear Research, Dubna, Russia}
	
\date{}
	
\begin{document}
	\maketitle

	\begin{abstract}
	Deuteron passage through a nonpolarized target is accompanied by birefringence effect which reveals as diverse phenomena, namely: rotation of spin and tensor polarization about the momentum direction, spin oscillations, vector polarization conversion to tensor that and vice versa, spin dichroism.
	Possibility to study deuteron spin dichroism effect i.e. the effect of tensor polarization acquiring by a nonpolarized deuteron beam moving in NuclotronM and passing through its internal target is discussed.  
\end{abstract}


\section{Introduction}

%

Investigation of nuclear reactions at interaction of polarized particles (nuclei)    with either an internal or an external target, as well as in the experiments with colliding polarized and nonpolarized beams are included into scientific programs of world-class research centers for particle physics.

It was shown in \cite{1992VG,1993VG} that
{quasi-optical birefringence} phenomenon arises 
for a particle (nuclei) beam with spin $S\geq1$ passing through nonpolarized matter.
This phenomenon  reveals itself as diverse effects, namely: spin  and tensor polarization  rotation around the momentum direction, spin oscillations, vector polarization conversion to tensor that and vice versa.
This phenomenon also exhibits spin dichroism.
%
%
%
{Spin dichroism phenomenon} leads to acquiring  tensor polarization for an initially nonpolarized particle beam with spin $S\geq1$, when the beam passes through a nonpolarized target.
In case, when a beam is polarized, vector polarization of the beam rotates and tensor polarization is converted  to vector that and vice versa, similar to optical birefringence in anisotropic media, when circular polarization of light converts linear that and vice versa.

The above mentioned phenomena are caused by the refractive index dependence  on the direction of particle  spin. 
For example, deuteron refractive index for spin projection $m=\pm1$ is not equal to that for $m=0$ (see~\cite{1992VG,1993VG}).

However, unlike photons, for particles with nonzero rest mass the birefringence effect exists in a homogeneous isotropic medium, even if the medium contains spinless or nonpolarized nuclei.
The point is that the effect is caused by an intrinsic anisotropy possessed  by the particles  with spin $S \ge 1$ themselves rather than medium anisotropy (unlike particles with spin $0$ and $1/2$).

%

%
The phenomenon of spin dichroism was first observed in joint experiment
at Universit\"{a}t zu K\"{o}ln (Germany) prepared by {teams from Institut f\"{u}r Kernphysik, Forschungszentrum J\"{u}lich; Institut f\"{u}r Kernphysik, Universit\"{a}t zu K\"{o}ln; Institute for Nuclear Problems, Minsk; PNPI, Gatchina} for deuterons in the energy range $5-20$ MeV
\cite{spinorb_ex1,spinorb_ex2,rins_63,VKB_source,Sey2011}
and at JINR (Dubna, Russia) for deuterons with momentum 5 GeV/c
 with the use of external target at Nuclotron
\cite{spinorb_ex3,Azhgirei1,Azhgirei}.

%
This article presents analysis of possibility to observe spin dichroism effect at Nuclotron internal target  at the Nuclotron M – NICA accelerator complex.

This paper is organized as follows. First, the general description of
birefringence phenomenon (spin oscillation and spin dichroism)
for particles with spin $ S\ge 1$ is provided.
Then, in section \ref{sec:evolution}  we consider evolution of polarization characteristics of a particle beam in an internal target of Nuclotron.
Dichroism effect for a deuteron beam moving in Nuclotron with internal target is evaluated in section \ref{sec:deuteron}.  
Acquiring tensor polarization for a particle beam in the Nuclotron ring is shown to be measurable with the existing detection system ~\cite{2011Kurilkin} based on CH$_2$ polarimeter.

\section{The phenomenon of birefringence (spin oscillation and spin dichroism)
	of particles with spin $ S\ge 1$\label{ch:eikanal3}}


The refractive index of particles with spin
$S \ge 1$ can be written \cite{1992VG,1993VG} as follows:
\begin{equation}
 \hat{N}=1+\frac{2\pi\rho}{k^{2}}\hat{f}(0)\,,
 \label{18.15}
\end{equation}
where $\hat{f}(0)=\textrm{Tr}\hat{\rho}_{J}\hat{F}(0)$;
$\hat{\rho}_{J}$ is the spin density matrix of the scatterer;
$\hat{F}(0)$ is the operator amplitude of forward scattering that
acts in the spin space of the particle and the scatterer with spin
$J$, $\rho$ is the number of atoms per cm$^3$. 
The explicit expression for the forward scattering amplitude $\hat{f}(0)$ in the non-relativistic approximation see in Appendix. 

If at entering the target the particle wave
function is $\psi_{0}$, then after passing the path length  $z$,  it will be
$\psi=\exp[ik\hat{N}z]\psi_{0}$.

Three parameters enable to describe forward
scattering: $\vec{S}$, $\vec{J}$, and $\vec{n}=\vec{k}/k$;
$\vec{k}$ is the particle wave vector.

It is known (see, for example, \cite{Varshalovich_ru,Varshalovich_en,VKB_Ohlsen,NO_RU,NO_EN}) that the spin matrix of dimensionality
$(2S+1)(2S+1)$ can be expanded in terms of a complete set of
$(2S+1)^{2}$ matrices, in particular, in terms of a set of
polarization operators $\hat{T}_{LM}(S)$, where $0\ll L\ll 2S$,
$-L\ll M\ll L$.
%

The most general form of such expansion allowing for the fact that
$\hat{F}$ should be scalar with respect to rotations is as follows \cite{1983VG,1983VG_en,1992VG,1993VG}:
\begin{eqnarray}
\hat{F}&=& A+A_{1}\hat{S_{i}}\hat{J_{i}}+A_{2}\hat{S_{i}}\hat{J_{k}}n_{i}n_{k}+A_{3}\hat{J_{i}}\hat{J_{k}}n_{i}n_{k}\nonumber\\
&+& A_{4}\hat{S_{i}}\hat{S_{k}}n_{i}n_{k}+A_{5}\hat{S_{i}}\hat{S_{k}}\hat{J_{i}}{J_{k}}+A_{6}\hat{S_{i}}\hat{S_{k}}n_{i}n_{k}\hat{J_{l}}\hat{J_{m}}n_{l}n_{m}+\ldots \nonumber\\
& & \ldots + B\hat{S_{i}}n_{i}+B_{1}\hat{S_{i}}\hat{J_{m}}e_{iml}n_{l}+B_{2}\hat{S_{i}}n_{i}\hat{S_{l}}{J_{l}}+B_{3}\hat{S_{i}}\hat{S_{l}}n_{i}n_{l}\hat{J_{m}}n_{m}\nonumber\\
&+&B_{4}\hat{J_{i}}n_{i}+B_{5}\hat{S_{i}}\hat{J_{m}}e_{iml}n_{l}\hat{S_{p}}n_{p}+\ldots,
\label{18.17}
\end{eqnarray}
where 
terms proportional to amplitudes A are P- and T-even;
those proportional to B, B$_2$, B$_3$, B$_4$ are P-odd and T-even;
one proportional to B$_1$ is P- and T-odd;
one proportional to B$_5$ is P-even and T-odd;
three dots stand for the terms containing the
products of $\hat{S_{i}}$ and $\hat{J_{i}}$ up to $2S$ and $2J$.

Upon averaging $\hat{F}$ using the spin density matrix of the
target nuclei, we find the explicit form of a coherent elastic
zero--angle scattering amplitude, and hence the refractive index
and the particle wave function in the target. According to
\ref{18.17}, for particles with spin $S>1/2$, there appear
additional terms involving spin operators in the second and higher
powers.

Let us find out what these terms lead to. We shall first pay
attention to the fact that even in the case of a nonpolarized
target, the amplitude $\hat{f}(0)$ depends on the spin operator of
the incident particle and, when the quantization axis $z$ is
directed along $\vec{n}$,  can be written in the form
\begin{equation}
 \hat{f}(0)=d+d_{1}\hat{S_{z}}^{2}+d_{2}\hat{S_{z}}^{4}\ldots
+d_{s}\hat{S_{z}}^{2s}\,.
\label{18.18}
\end{equation}

We consider a
specific case of strong interactions, invariant with respect to
time and space reflections; for this reason, the terms containing the
odd powers of $\hat{S}$ are dropped. According to (\ref{18.15}), the refractive index
is
\begin{equation}
\hat{N}=1+\frac{2\pi\rho}{k^{2}}(d+d_{1}\hat{S_{z}}^{2}+d_{2}\hat{S_{z}}^{4}\ldots
+d_{s}\hat{S_{z}}^{2s})
\label{18.19}
\end{equation}
that yields an important conclusion, namely: dependence of the refractive index of a particle with spin $S>1/2$  on the spin orientation with respect to the momentum direction. 
Write $m$ for
a magnetic quantum number, then the refractive index of a particle
in the state which is the eigenstate of the operator $\hat{S_{z}}$ of
the spin projection on the $z$-axis is
\begin{equation}
N(m)=1+\frac{2\pi\rho}{k^{2}}(d+d_{1}m^{2}+d_{2}m^{4}+ \ldots
+d_{s}m^{2s})\,.
\label{18.20}
\end{equation}
According to (\ref{18.20}), the states of a particle with quantum
numbers $m$ and ($-m$) have the same refractive indices. For a
particle with spin $1$ (for example, a $J/ \psi$-particle,
deuteron)  and for a  particle with spin $3/2$ (for example, Ne$^{21}$ nucleus)
\begin{equation}
 N(m)=1+\frac{2\pi\rho}{k^{2}}(d+d_{1}m^{2})\,.
 \label{18.21}
\end{equation}
As is seen, $\texttt{Re} N(\pm 1)\neq \texttt{Re} N(0)$;
$\texttt{Im} N(\pm 1)\neq \texttt{Im} n(0)$; $\texttt{Re} N(\pm
3/2)\neq \texttt{Re} N(\pm1/2)$; $\texttt{Im} N(\pm 3/2)\neq
\texttt{Im} N(\pm 1/2)$.

From this follows that for particles with spin $S>1/2$, even a
nonpolarized target causes spin dichroism: due to different
absorption, the initially nonpolarized beam passing through matter
acquires polarization, or  more precisely, alignment
\cite{1992VG,1993VG}.

In view of the above analysis, from
(\ref{18.19})--(\ref{18.21}) follows that in a medium, a moving
particle with spin $S \ge 1$ possesses a potential energy:
\begin{eqnarray*}
	\hat{V}=-\frac{2\pi \hbar^{2} \rho}{M}(d+d_{1} \hat{S_{z}}^{2}+d_{2}\hat{S_{z}}^{4}+\ldots)\,,\\
	V(m)=-\frac{2\pi \hbar^{2}
		\rho}{M}(d+d_{1}m^{2}+d_{2}m^{4}+\ldots)\,.
\end{eqnarray*}
The expression for  $\hat{V}$, which describes interaction between
the particle and matter  is similar to that between the atom of
spin $S\ge 1$ and the electric field.
As a result, the spin levels of the particle in matter split in
a way similar to Stark splitting of atomic levels in the electric field.
Hence, we may say that a particle of spin $S \ge 1$, moving in matter,
experiences the influence of a certain pseudoelectric field
(compare with the introduction of a pseudomagnetic field).

Since we have obtained the explicit spin structure of the
refractive index, then we know the wave function $\psi$, and for
every particular case we can find all spin characteristics of the
beam, which passed distance $z$in a target.

\subsection
{Rotation and Oscillation of Deuteron Spin \\in Nonpolarized
	Matter and Spin Dichroism \\(Birefringence Phenomenon)}
\label{cosy_sec:1.1}

We shall further dwell on the passage of deuterons through matter.

According to (\ref{18.21}), the refractive indices for the states
with ${m=+1}$ and ${m=-1}$ are the same, while those for for the
states with $m=\pm 1$ and $m=0$ are different
($\texttt{Re}{\textit{N}(\pm 1)} \neq \texttt{Re}{\textit{N}(0)}$
and $\texttt{Im}{\textit{N}(\pm 1)} \neq
\texttt{Im}{\textit{N}(0)}$).

\begin{figure}[htbp]
	\epsfxsize = 10 cm \centerline{\epsfbox{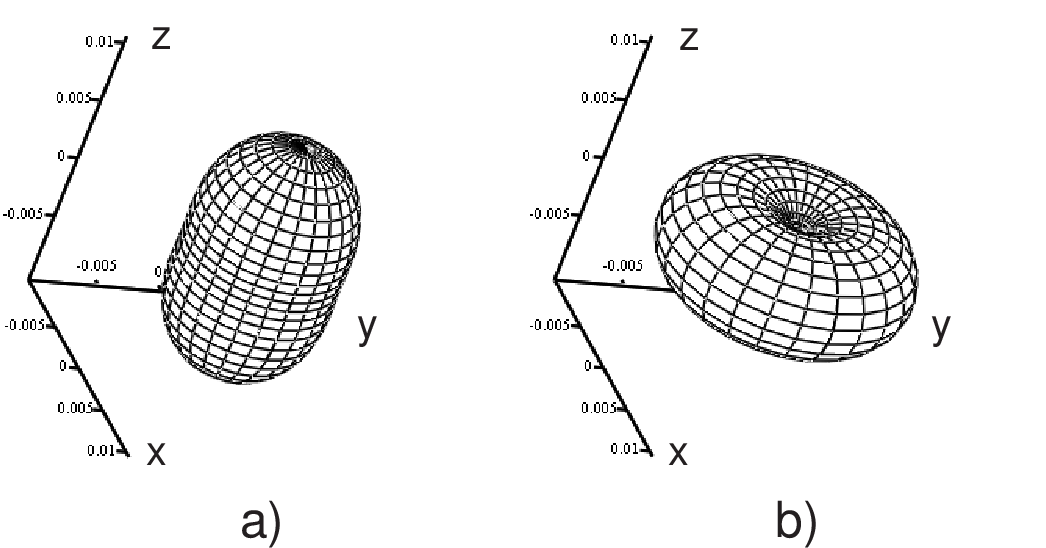}}
	\caption{Squared module for deuteron ground state wave functions
		for the distance of 1.8 fm between its nucleons in the states a)
		$m=\pm 1$; b) $m=0$ }
	\label{VKB_fig1}
\end{figure}

\begin{figure}[htbp]
	\epsfxsize =8 cm \centerline{\epsfbox{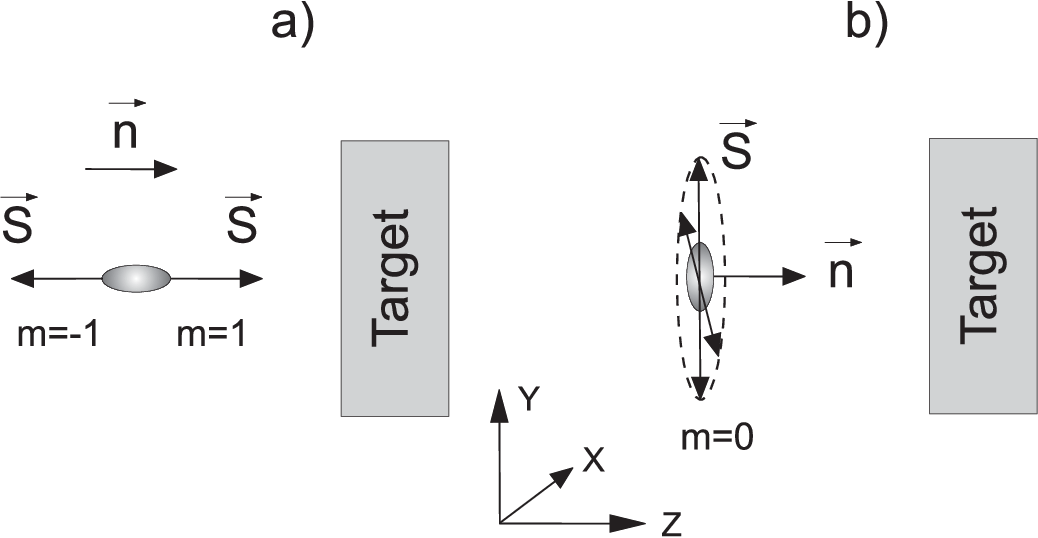}}
	\caption{Two possible orientation of vectors $\vec {S}$ and $\vec
		{n}=\frac{\vec{k}}{k}$: a) $m=\pm 1$; b) $m=0$}
	\label{cosy_sigma}
\end{figure}

This can be explained as follows (see Fig. \ref{VKB_fig1}, Fig. \ref{cosy_sigma}):
the shape of a deuteron in the ground state is non--spherical.
Therefore, the scattering cross section $\sigma_{\pm 1}$ for a
deuteron with $m= \pm 1$ (deuteron spin is parallel (antiparallel) to its
momentum $\vec{k}$) differs from the scattering cross section
$\sigma_{0}$ for a deuteron with $m=0$:
\begin{equation}
 \sigma_{\pm 1} \ne \sigma_{0} ~\Rightarrow ~
\texttt{Im}\textit{f}_{\pm 1}(0)=\frac{k}{4\pi} \sigma_{\pm 1}\neq
\texttt{Im}\textit{f}_ 0 (0)=\frac{k}{4\pi}\sigma_0\,.
\label{cosy_eq7}
\end{equation}
According to the dispersion relation, $\texttt{Re}{\textit{f}(0)}
\sim\Phi (\texttt{Im}{\textit{f}(0))}$, hence
${\texttt{Re}{\textit{f}_{0}(0)} \neq \texttt{Re}{\textit{f}_{\pm
			1}(0)}}$

From the above follows that deuteron spin dichroism appears even
when a deuteron passes through a nonpolarized target: owing to
the fact that beam absorption depends on the orientation of the
deuteron spin, the initially nonpolarized beam acquires alignment.

Let us consider the deuteron spin state in a target.
The spin state of the deuteron is described by its vector and
tensor  polarizations $\vec {p} = \langle \hat{\vec {S}}\rangle$ and
$p_{ik} =
\langle \hat{Q}_{ik} \rangle$, respectively.
As the deuteron  moves in matter, its vector and tensor
polarizations  change.
To calculate $\vec {p}$ and $p_{ik}$, one needs to know the explicit
form of the deuteron spin wave function $\psi$.

The wave function of the deuteron that has passed the distance $z$
inside the target is:
\begin{equation}
\psi \left( {z} \right) = 
{\texttt{e}}^{ik\hat{N}z} \psi _{0}\,,
\label{cosy_psiz}
\end{equation}
where $\psi_{0}$ is the wave function of the deuteron before
entering the target.
The wave function $\psi$ can be expressed as a superposition of
the basic spin functions $\chi_{m}$, which are the eigenfunctions
of the operators $\hat{S}^{2}$ and $\hat{S}_{z}$ ($\hat {S}_{z}
\chi _{m} = m\chi _{m}$):
\begin{equation}
 \psi = \sum\limits_{m = \pm 1,0} {a^{m}\chi
	_{m}}\,.
\label{cosy_psi}
\end{equation}
Therefore
\begin{equation}
\begin{array}{l}
\Psi = \left( \begin{array}{*{20}c}
{a^{1}} \hfill \\
{a^{0}} \hfill \\
{a^{ - 1}} \hfill \\
\end{array}  \right) = \left( {{\begin{array}{*{20}c}
		{a\,{\texttt{e}}^{i\delta _{1}}\, {\texttt{e}}^{ikN_{+1} z}} \hfill \\
		{b\,{\texttt{e}}^{i\delta _{0}}\, {\texttt{e}}^{ikN_{0} z}} \hfill \\
		{c\,{\texttt{e}}^{i\delta _{ - 1}}\, {\texttt{e}}^{ikN_{ - 1} z}} \hfill \\
		\end{array}} } \right) =  
	\left( {{\begin{array}{*{20}c}
		{a\,{\texttt{e}}^{i\delta _{+1}}\, {\texttt{e}}^{ikN_{+1} z}} \hfill \\
		{b\,{\texttt{e}}^{i\delta _{0}}\, {\texttt{e}}^{ikN_{0} z}} \hfill \\
		{c\,{\texttt{e}}^{i\delta _{ - 1}}\, {\texttt{e}}^{ikN_{1} z}} \hfill \\
		\end{array}}} \right)\,,
\end{array}
\label{cosy_psidepth}
\end{equation}
here equality $N_{1}=N_{-1}$ is used.

Suppose that the plane $(yz)$ coincides with the plane formed by the
initial vector polarization $\vec {p}_0 \neq 0$ and the
momentum $\vec{k}$ of the deuteron. In this case
\[
\delta
_{+1}-\delta _{0}= \delta _{0}-\delta _{-1}=\frac{{\pi}}{{2}},
\]
and the components of the polarization vector at $z = 0$ are $p_x =
0,p_y\neq 0$, and $p_z\neq 0$.

The components of the vector polarization are defined as:
\[
\vec{p}=\langle \hat{\vec
{S}}
\rangle = \frac{{\langle \Psi \left| \hat{\vec {S}} \right|\Psi
		\rangle} }{{\left\langle {{\Psi} } \mathrel{\left| {\vphantom
				{{\Psi} {\Psi} }} \right. \kern-\nulldelimiterspace} {{\Psi} }
		\right\rangle} }
\]
inside the target they can be expressed as follows:
\begin{eqnarray}
	\label{cosy_rot1}
p_x&=& \frac{{ \sqrt {2} \,{\texttt{e}}^{ - \frac{{1}}{{2}}\rho \left( {\sigma
				_{0} + \sigma _{1}} \right)z}b\left( {a - c} \right)\sin\left(
		{\frac{{2\pi \rho} }{{k}}\texttt{Re} d_{1} z}
		\right)}}{{\left\langle {{\Psi} } \mathrel{\left| {\vphantom
				{{\Psi}  {\Psi} }} \right. \kern-\nulldelimiterspace}
		{{\Psi} } \right\rangle} } ,\nonumber\\
p_y&=&\frac{{\sqrt {2} \,{\texttt{e}}^{ - \frac{{1}}{{2}}\rho \left( {\sigma
				_{0} + \sigma _{1}} \right)z}b\left( {a + c} \right)\cos\left(
		{\frac{{2\pi \rho} }{{k}}\texttt{Re} d_{1} z}
		\right)}}{{\left\langle {{\Psi} } \mathrel{\left| {\vphantom
				{{\Psi}  {\Psi} }} \right. \kern-\nulldelimiterspace}
		{{\Psi} } \right\rangle} } ,\nonumber \\
p_z& =&\frac{{\,{\texttt{e}}^{\rho \sigma _{1} z}\left( {a^{2} - c^{2}}
		\right)}}{{\left\langle {{\Psi} } \mathrel{\left| {\vphantom
				{{\Psi}  {\Psi }}} \right. \kern-\nulldelimiterspace} {{\Psi} }
		\right\rangle} }\,.
\\ \nonumber
\end{eqnarray}
Similarly, the components of the tensor polarization
\[
\hat
{Q}_{ij} = \frac{{3}}{{2}}\left( {\hat {S}_{i} \hat {S}_{j} + \hat
	{S}_{j} \hat {S}_{i} - \frac{{4}}{{3}}\delta _{ij}} \right)
\]
are expressed as
\[
p_{ik}=\langle \hat
{Q}_{ij}\rangle = \frac{{\langle \Psi \left| \hat
		{Q}_{ij} \right|\Psi
		\rangle} }{{\left\langle {{\Psi} } \mathrel{\left| {\vphantom
				{{\Psi} {\Psi} }} \right. \kern-\nulldelimiterspace} {{\Psi} }
		\right\rangle} }
\]
and read as follows:
\\
\begin{eqnarray}
	\label{cosy_rot2}
p_{xx}&=&\frac{{ - \frac{{1}}{{2}}\left( {a^{2} + c^{2}}
		\right)\,{\texttt{e}}^{ - \rho \sigma _{1} z} + b^{2}\,{\texttt{e}}^{ - \rho \sigma _{0} z}
		- 3ac\,{\texttt{e}}^{ - \rho \sigma _{1} z}}}{{\left\langle {{\Psi} }
		\mathrel{\left| {\vphantom {{\Psi}  {\Psi} }} \right.
			\kern-\nulldelimiterspace} {{\Psi} } \right\rangle} }
\,,\nonumber
\\
p_{yy}&=&\frac{{ - \frac{{1}}{{2}}\left( {a^{2} + c^{2}}
		\right)\,{\texttt{e}}^{ - \rho \sigma _{1} z} + b^{2}\,{\texttt{e}}^{ - \rho \sigma _{0} z}
		+ 3ac\,{\texttt{e}}^{ - \rho \sigma _{1} z}}}{{\left\langle {{\Psi} }
		\mathrel{\left| {\vphantom {{\Psi}  {\Psi} }} \right.
			\kern-\nulldelimiterspace} {{\Psi} } \right\rangle} }  \,,
\nonumber
\\
p_{zz}&=&\frac{{\left( {a^{2} + c^{2}}
		\right)\,{\texttt{e}}^{ - \rho \sigma _{1} z} - 2b^{2}\,{\texttt{e}}^{ - \rho \sigma _{0}
			z}}}{{\left\langle {{\Psi} } \mathrel{\left| {\vphantom {{\Psi}
					{\Psi} }} \right. \kern-\nulldelimiterspace} {{\Psi} }
		\right\rangle} } ,\nonumber\\
p_{xy}&=& 0 \,,
\\
p_{xz}&=&\frac{{ \frac{{3}}{{\sqrt {2}} }\,{\texttt{e}}^{ - \frac{{1}}{{2}}\rho
			\left( {\sigma _{0} + \sigma _{1}} \right)z}b\left( {a + c}
		\right)\sin\left( {\frac{{2\pi \rho} }{{k}}\texttt{Re} d_{1} z}
		\right)}}{{\left\langle {{\Psi} } \mathrel{\left| {\vphantom
				{{\Psi}  {\Psi }}} \right. \kern-\nulldelimiterspace} {{\Psi} }
		\right\rangle} }  \,,\nonumber
\\
p_{yz}&=&\frac{{\frac{{3}}{{\sqrt {2}} }\,{\texttt{e}}^{ - \frac{{1}}{{2}}\rho
			\left( {\sigma _{0} + \sigma _{1}} \right)z}b\left( {a - c}
		\right)\cos\left( {\frac{{2\pi \rho} }{{k}}\texttt{Re} d_{1} z}
		\right)}}{{\left\langle {{\Psi} } \mathrel{\left| {\vphantom
				{{\Psi}  {\Psi }}} \right. \kern-\nulldelimiterspace}
		{{\Psi} } \right\rangle} } \,,\nonumber\\
p_{xx}&+&p_{yy}+p_{zz}=0\,, \nonumber 
\end{eqnarray}
\\
\noindent  where
\begin{eqnarray*}
\left\langle {{\Psi} } \mathrel{\left| {\vphantom {{\Psi}
			{\Psi} }} \right. \kern-\nulldelimiterspace} {{\Psi} }
\right\rangle &=& \left( {a^{2} + c^{2}} \right)\,{\texttt{e}}^{ - \rho \sigma
	_{1} z} + b^{2}\,{\texttt{e}}^{ - \rho \sigma _{0} z},\nonumber\\
\sigma_{0}&=&\frac{{4\pi} }{{k}}\texttt{Im} f_0,\, \sigma _{1} = \frac{{4\pi}}{{k}}\texttt{Im} f_1,\\
f_0&=&d,\, f_1=d+d_1\,.\nonumber
\label{cosy_rot2+}
\end{eqnarray*}

According to \eqref{cosy_rot1}, \eqref{cosy_rot2} spin rotation and oscillation occur when
the angle between the polarization vector $\vec{p}$ and the momentum
$\vec{k}$ of the particle differs from $\pi/2$. The magnitude of the effect is determined by the phase
\begin{equation}
\label{cosy_ins} \varphi=\frac{2\pi\rho}{k}\texttt{Re} d_1 z\,.
\end{equation}

For example, let  $\texttt{Re} d_1 >0$. If the angle between the
polarization vector and momentum is acute, then the spin rotates
anticlockwise about the momentum direction, whereas the obtuse
angle between the polarization vector and the momentum gives rise
to a clockwise  spin rotation.

When the polarization vector and momentum are perpendicular
(transversely polarized particle), the components of
the vector polarization at $z = 0$ are: $p_x= 0$, $p_y\ne0$, and $p_z=0$.
In this case $a=c$ and the dependence of the vector polarization
on $z$ can be expressed as:
\begin{eqnarray}
		\label{cosy_ins11}
p_x&=&0,\nonumber\\
p_y&=&\frac{{\sqrt {2} \,{\texttt{e}}^{ - \frac{{1}}{{2}}\rho \left( {\sigma
				_{0} + \sigma _{1}} \right)z}2ba\cos\left( {\frac{{2\pi \rho
			}}{{k}}\texttt{Re} d_{1} z} \right)}}{{\left\langle {{\Psi} }
		\mathrel{\left| {\vphantom {{\Psi}  {\Psi} }} \right.
			\kern-\nulldelimiterspace} {{\Psi} } \right\rangle} }\,,\nonumber \\
p_z&=&0\,,\nonumber \\
p_{xx}&=&\frac{{ - 4a^{2}\,{\texttt{e}}^{ - \rho \sigma _{1} z}
		+ b^{2}\,{\texttt{e}}^{ - \rho \sigma _{0} z}}}{{\left\langle {{\Psi} }
		\mathrel{\left| {\vphantom {{\Psi}  {\Psi} }} \right.
			\kern-\nulldelimiterspace} {{\Psi} } \right\rangle} }\,, \\
p_{yy}&=&\frac{{2a^{2}\,{\texttt{e}}^{ - \rho \sigma _{1} z} + b^{2}\,{\texttt{e}}^{ - \rho
			\sigma _{0} z}}}{{\left\langle {{\Psi} } \mathrel{\left|
			{\vphantom {{\Psi}  {\Psi} }} \right.
			\kern-\nulldelimiterspace} {{\Psi} } \right\rangle} }\,,\nonumber\\
p_{zz}&=&\frac{{2a^{2}\,{\texttt{e}}^{ - \rho \sigma_{1} z} - 2b^{2}\,{\texttt{e}}^{ - \rho
			\sigma _{0} z}}}{{\left\langle {{\Psi} } \mathrel{\left|
			{\vphantom {{\Psi}  {\Psi} }} \right. \kern-\nulldelimiterspace}
		{{\Psi} } \right\rangle} }\,, \nonumber \\
p_{xz}&=&\frac{{  \frac{{3}}{{\sqrt {2}} }\,{\texttt{e}}^{ -
			\frac{{1}}{{2}}\rho \left( {\sigma _{0} + \sigma _{1}}
			\right)z}2ab\sin\left( {\frac{{2\pi \rho} }{{k}}\texttt{Re} d_{1}
			z} \right)}}{{\left\langle {{\Psi} } \mathrel{\left| {\vphantom
				{{\Psi}  {\Psi }}} \right. \kern-\nulldelimiterspace} {{\Psi} }
		\right\rangle} }\,,  \nonumber \\
	p_{xz}&=&0\,, \nonumber\\
	p_{xx}&+&p_{yy}+p_{zz}=0\,. \nonumber
\end{eqnarray}
According to (\ref{cosy_ins11}), no rotation occurs in this case; the vector and tensor polarization
oscillate when a transversely polarized deuteron passes through
matter.

\subsection
{The Effect of Tensor Polarization Emerging in Nonpolarized Beams
	Moving in Nonpolarized Matter (birefringence spin dichroism effect)} \label{cosy_sec:1.2}

The birefringence effect, in particular,
acquiring tensor polarization by an initially nonpolarized beam passing through a nonpolarized target,
can be most clearly described using the spin density matrix \cite{spinorb_ex2}.

For deuterons (spin $S=1)$, the spin density matrix for the  beam before entering the target can be
written as follows:
\begin{equation}
\hat{\rho}_{0}=\frac{1}{3}\hat{\texttt{I}}+\frac{1}{2}\vec{p}_{0}\hat{\vec{S}}+\frac{1}{9}p^{(0)}_{ik}\hat{Q}_{ik}\,,
\label{cosy_ins+}
\end{equation}
where $\hat{\texttt{I}}$ is the identity (unit) matrix, $\vec{p}_{0}$ is the polarization vector of the beam,
$p^{(0)}_{ik}$  is the polarization vector of the beam incident on the target.
Using (\ref{cosy_psiz}), one can express the density matrix of the
deuteron beam in the target as:
\begin{equation}
\hat {\rho}  = \,{\texttt{e}}^{ik\hat{N}z}\hat{\rho}_{0}\,{\texttt{e}}^{-ik\hat{N}^{\ast}z}\,.
\label{cosy_ins+1}
\end{equation}
As a result, we have
\begin{equation}
\vec{p}=\langle \hat{\vec{ S}}\rangle = \frac{{{\textrm{Tr}} \left( \hat
		{\rho} \hat{\vec{ S}}\right)}}{{ \textrm{Tr}\hat {\rho}		}},\quad p_{ik}= \langle \hat{Q}_{ik} \rangle =
\frac{{\textrm{Tr}} \left( \hat {\rho}\hat{Q}_{ik} \right)}{
	\textrm{Tr} \hat {\rho} }\,,
\label{cosy_ins+2}
\end{equation}
where $i,k=x,y,z$.

In case of thin targets, the vector and tensor polarization of
the deuteron beam inside the target  with the use of
the first-order approximation for exponent $\,{\texttt{e}}^{ik( \hat{N}-1)z} \approx
1+ik(\hat{N}-1 )z$ can be expressed as follows:
\begin{eqnarray}
\label{cosy_polpar1}
p_{x}&=& \frac{{\left[ {1 - \frac{{1}}{{2}}\rho z\left( {\sigma
				_{0} + \sigma _{1}}  \right)} \right] p_{x,0}+
		\frac{{4}}{{3}}\frac{{\pi \rho z}}{k}\texttt{Re} d_{1}
		p_{yz}^{(0)}}}{{\textrm{Tr}\hat {\rho }}}\,,\nonumber
\\
p_{y} &=& \frac{{\left[ {1 - \frac{{1}}{{2}}\rho z\left( {\sigma
				_{0} + \sigma _{1}}  \right)} \right] p_{y,0}  -
		\frac{{4}}{{3}}\frac{{\pi \rho z}}{k} \texttt{Re} d_{1}
		p_{xz}^{(0)}}}{{\textrm{Tr}\hat {\rho }}}\,, \nonumber
\\
p_{z}& =& \frac{{\left( {1 - \rho \sigma _{1} z} \right)p_{z,0}} }
{{\textrm{Tr}\hat {\rho} }}\,,\nonumber
\\
p_{xx}&=& \frac{{\left( {1 - \rho \sigma _{1} z} \right)p_{xx}^{(0)} +
		\frac{{1}}{{3}}\rho z\left( {\sigma _{1} - \sigma _{0}} \right) -
		\frac{{1}}{{3}}\rho z\left( {\sigma _{1} - \sigma _{0}}
		\right)p_{zz}^{(0)}} }{{\textrm{Tr}\hat {\rho} }}\,, \nonumber
\\
p_{yy} &=& \frac{{\left( {1 - \rho \sigma _{1} z} \right)p_{yy}^{(0)}
		+ \frac{{1}}{{3}}\rho z\left( {\sigma _{1} - \sigma _{0}}  \right)
		- \frac{{1}}{{3}}\rho z\left( {\sigma _{1} - \sigma _{0}}
		\right)p_{zz}^{(0)}} }{{\textrm{Tr}\hat {\rho} }}\,, \nonumber
\\
p_{zz} &=& \frac{{\left[ {1 - \frac{{1}}{{3}}\rho z\left( {2\sigma
				_{0} + \sigma _{1}}  \right)} \right]p_{zz}^{(0)} -
		\frac{{2}}{{3}}\rho z\left( {\sigma _{1} - \sigma _{0}}
		\right)}}{{\textrm{Tr}\hat {\rho} }}\,, \nonumber
\\
p_{xy} &=& \frac{{\left( {1 - \rho \sigma _{1} z} \right)p_{xy}^{(0)}}
} {{\textrm{Tr}\hat {\rho} }}\,,
\\
p_{xz} &=& \frac{{\left[ {1 - \frac{{1}}{{2}}\rho z\left( {\sigma
				_{0} + \sigma _{1}}  \right)} \right] p_{xz}^{(0)} + 3\frac{{\pi \rho
				z}}{{k}}\texttt{Re} d_{1} p_{y,0} } }{{\textrm{Tr}\hat {\rho }}}\,,\nonumber
\\
p_{yz} &=& \frac{{\left[ {1 - \frac{{1}}{{2}}\rho z\left( {\sigma
				_{0} + \sigma _{1}} \right)} \right]p_{yz}^{(0)} - 3\frac{{\pi \rho
				z}}{{k}}\texttt{Re} d_{1} p_{x,0}} }{{\textrm{Tr}\hat {\rho }}}\,, \nonumber
\end{eqnarray}
\noindent where
\[
\textrm{Tr}\hat {\rho} = 1 - \frac{{\rho z}}{{3}}\left(
{2\sigma _{1} + \sigma _{0}}  \right) - \frac{{\rho z}}{{3}}\left(
{\sigma _{1} - \sigma _{0}} \right)p_{zz}^{(0)}\, .
\]

In accordance with \eqref{cosy_polpar1}, if the initial beam polarization is zero $p_{x,0} = p_{y,0} = p_{z,0} = 0$ in contrast to the tensor polarization components, which are nonzero $p_{xz}^{(0)} = p_{yz}^{(0)} \ne 0$, beam motion in a target is accompanied by appearance of nonzero vector polarization  components $p_{x}$ or  $p_{y}$.
In case if a beam initially  has no tensor  polarization, while vector polarization  components $p_{x,0} = p_{y,0} \ne 0$, beam motion in the target leads to appearance of nonzero tensor polarization components: vector component $p_x$ converts to tensor component $p_{yz}$, vector component $p_y$ converts to tensor component $p_{xz}$ and vice versa tensor components $p_{yz}$ and $p_{xz}$ convert to vector components $p_x$ and $p_y$, respectively.

If the beam is initially nonpolarized ($p_{x,0} = p_{y,0} =
p_{z,0} = p_{xx}^{(0)} = p_{yy}^{(0)} =p_{zz}^{(0)} = $ $p_{xy}^{(0)} = p_{xz}^{(0)} =
p_{yz}^{(0)} = 0$), then after passing through the nonpolarized target
of thickness $z$, the deuteron beam acquires  tensor polarization:
\begin{eqnarray}
	\label{eq:pzzext}
p_{zz}&\approx& - \frac{{2}}{{3}}\rho z\left( {\sigma _{\pm 1} -
	\sigma _{0}}  \right)= \frac{{2}}{{3}}\rho z \,\Delta \sigma \, ,\nonumber\\
p_{xx} &=& p_{yy} \approx \frac{{1}}{{3}}\rho z\left( {\sigma _{\pm 1}
	- \sigma _{0}}  \right)= - \frac{{1}}{{3}}\rho z \,\Delta \sigma\,, \label{cosy_polpar2}
\end{eqnarray}
where $\Delta \sigma = \sigma _{0} - \sigma _{\pm 1} $, $ \sigma_{+1}= \sigma _{-1}$.
Vector polarization remains equal to zero.


%
The expression for tensor polarization (\ref{cosy_polpar2}) may also be obtained from another viewpoint.

Let a deuteron beam in spin state with $m=+1$ pass through a
target. The beam intensity changes as
$I_{+1}(z)=I_{+1}^0\,{\texttt{e}}^{-\sigma_{+1}\rho z}$, where $I_{+1}^0$ is the beam
intensity before entering the target. Similarly, for states with
$m={-1}$ and $m=0$, the intensity changes as
$I_{-1}(z)=I_{-1}^0\,{\texttt{e}}^{-\sigma_{-1}\rho z}$ and
$I_0(z)=I_{0}^0\,{\texttt{e}}^{-\sigma_0\rho z}$, where $I_{-1}^0$ and
$I_{0}^0$ are the beam intensities before entering the target,
respectively.

Let us consider the transmission of a nonpolarized deuteron beam
through a nonpolarized target.
The nonpolarized deuteron beam can be described as a composition
of three polarized beams with the equal intensities
$I=I_{+1}^0+I_{-1}^0+I_{0}^0$, {$I_{\pm1}^0=I_{0}^0=I/3$}.
In a real experiment
$\sigma_{\pm1(0)}\rho z\ll1$ and the change in the intensity for each
beam can be expressed as
$I_{\pm1}(z)=I_{\pm1}^0(1-\sigma_{\pm1}\rho z)$ and
$I_{0}(z)=I_{0}^0(1-\sigma_{0}\rho z)$.
According to \cite{VKB_Ohlsen}, the tensor polarization of the beam can
be expressed as
\[
p_{zz}=\frac{I_{-1}+I_{+1}-2I_0}{I_{-1}+I_{+1}+I_0}\,.
\]

The tensor polarization acquired due to spin dichroism effect by the initially nonpolarized deuteron
beam transmitting through the target reads as follows:
%
\begin{eqnarray}
\label{VKB_pzzdef}
p_{zz}(L)=\frac{I_{-1}(L)+I_{+1}(L)-2I_{0}(L)}{I_{-1}(L)+I_{+1}(L)+I_{0}(L)}
\approx \frac{2N_aL\left(\sigma_{0}-\sigma_{\pm1}\right)}{3M_r}=-\frac{8\pi
	N_aL \, \texttt{Im}(d_1)}{3kM_r}\,, 
\end{eqnarray}
where $N_a$ is the Avogadro number, $L$ is the target thickness in
g/cm$^2$, $M_r$ is the molar mass of the target matter.

%
Note that a deuteron passing through a target loses energy
by ionization of matter, then, taking into account the energy
change, we can write the tensor polarization as

\begin{eqnarray}
\label{VKB_pzzrot}
p_{zz}(L)&=&\frac{2N_a}{3M_r}\int_0^L
\left(\sigma_{0}\left(E\left(L'\right)\right)-\sigma_{\pm1}\left(E\left(L'\right)\right)\right)dL'\nonumber\\
&=&-\frac{8\pi N_a}{3M_r}
\int_0^L\frac{\texttt{Im}(d_1\left(E\left(L'\right)\right))}{k(L')}dL'\,.
\end{eqnarray}
According to (\ref{VKB_pzzrot}), the imaginary part of the
spin-dependent forward scattering amplitude can be measured
directly in a transmission experiment
by means of deuteron beam tensor polarization, which arises due to
deuteron spin dichroism.

Thus, theoretical studies of the deuteron beam transmission
through the nonpolarized target predict the appearance of tensor
polarization in a transmitted beam due to deuteron spin dichroism.

\subsection{Equations enabling to describe polarization vector and quadrupolarization tensor for a deuteron beam moving in Nuclotron with internal target}

Let us consider deuteron beam motion in a storage ring in the presence of external electric and magnetic fields.
The spin precession of the particle, caused by the interaction of the particle magnetic moment with the external electromagnetic field, is described by the Bargmann-Michel-Telegdi equation \cite{6,8}
\begin{equation}
	\frac{d\vec{p}}{dt}=[\vec{p}\times\vec{\Omega}_{0}],
	\label{2.1}
\end{equation}
where $t$ is the time in the laboratory frame,
\begin{equation}
	\vec{\Omega}_{0}=\frac{e}{mc}\left[\left(a+\frac{1}{\gamma}\right)\vec{B}
	-a\frac{\gamma}{\gamma+1}\left(\vec{\beta}\cdot\vec{B}\right)\vec{\beta}
	\right]
	,
	\label{2.2}
\end{equation}
$m$ is the particle mass, $e$ is its charge, $\vec{p}$ is the polarization vector, $\gamma$ is the Lorentz factor,
$\vec{\beta}=\vec{v}/c$, $\vec{v}$ is the particle velocity, $a=(g-2)/2$, $g$ is the gyromagnetic ratio,
$\vec{B}$ is the magnetic field at the particle location.

Thus, evolution of the deuteron spin is described by the following equation:
\begin{eqnarray}
\frac{d\vec{p}}{dt}=
\frac{e}{mc}\left[\vec{p}\times\left\{\left(a+\frac{1}{\gamma}\right)\vec{B}
-a\frac{\gamma}{\gamma+1}\left(\vec{\beta}\cdot\vec{B}\right)\vec{\beta}
\right\}
\right].
\label{2.4}
\end{eqnarray}

However, the equation (\ref{2.4}) alone is not sufficient to describe spin evolution in the Nuclotron with an internal target: it is necessary to supplement it with a contribution
caused by interaction of the deuteron with the internal target.
This interaction is described by effective potential energy $\hat{V}$, which a particle in matter possesses
\cite{Goldberger}:
\begin{equation}
\hat{V} =- \frac{2 \pi {\hbar}^2}{M} {\rho} \hat{f(0)},
\label{U1}
\end{equation}
where $\hat{f}(0)$ is the amplitude of elastic coherent forward scattering, the explicit expression of $\hat{f}(0)$ for a particle with  spin $S=1$  was obtained in   \cite{1992VG,1993VG,4}:
\begin{equation}
\hat{f(0)}=d+d_{1}\left(\hat{\vec{S}}\vec{n}\right)^{2},
\label{1.2}
\end{equation}
where  $\vec{n}$ is the unit vector in the direction of particle momentum.

The density matrix of a system ''deuteron beam + target''  can be expressed as:
\begin{eqnarray}
\hat{\rho}=\hat{\rho}_{d}\otimes \hat{\rho}_{t},
 \label{2.6}
\end{eqnarray}
where $\hat{\rho}_{d}$ is the density matrix of a deuteron beam, $\hat{\rho}_{t}$ density matrix of a target.
\noindent The density matrix of a deuteron beam
\begin{eqnarray}
\hat{\rho}_{d}=I(\vec{k})\left(\frac{1}{3}\hat{\texttt{I}}
+\frac{1}{2}\vec{p}(\vec{k})\hat{\vec{S}}+\frac{1}{9}p_{ik}(\vec{k})\hat{Q}_{ik}\right),
\label{2.7}
\end{eqnarray}
$I(\vec{k})$ is the beam intensity, $\vec{p}$ is the polarization vector, $p_{ik}$ is the polarization tensor for the deuteron beam.
For a nonpolarized target
 $\hat{\rho}_{t}=\hat{\texttt{I}}/3$, where
$\hat{\texttt{I}}$ is the unit matrix in the spin space of target particles (atoms, nuclei).

Equation for density matrix of a deuteron beam reads as: 
\begin{eqnarray}
\frac{d\hat{\rho}_{d}}{dt}=-\frac{i}{\hbar}\left[\hat{H},\hat{\rho}_{d}\right]{+\left(\frac{\partial\hat{\rho}_{d}}{\partial
t}\right)_{col}},
\label{2.9}
\end{eqnarray}
where $\hat{H}=\hat{H}_{0}+ {\hat{V}}$.
The term responsible for collisions   
$\left(\frac{\partial\hat{\rho}_{d}}{\partial t}\right)_{col}$ can be obtained with the use of method described in \cite{9}:
\begin{eqnarray}
\left(\frac{\partial\hat{\rho}_{d}}{\partial
t}\right)_{col}
=vN\emph{Sp}_{t}\left[\frac{2\pi
i}{k}\left[\hat{F}(\theta=0)\hat{\rho}-\hat{\rho} \hat{F}^{+}(\theta=0)\right] +\int
d\Omega \hat{F}(\vec{k}^{'})\hat{\rho}(\vec{k}^{'})\hat{F}^{+}(\vec{k}^{'})\right],
\label{2.11}
\end{eqnarray}
where $\vec{k}^{'}=\vec{k}+\vec{q}$, $\vec{q}$ is the momentum, transferred from the incident particle  to matter,
 $v$ is the velocity of the incident particle, $N$ is the number of atoms in cm$^3$ of matter,
 $\hat{F}$ is the scattering amplitude, which depends on spin operators of deuterons and nuclei (atoms) of matter, $\hat{F}^+$ is the operator Hermitian conjugate to operator $\hat{F}$.
The first term in (\ref{2.11}) describes coherent scattering of the particle by the nuclei of matter, while the second one is responsible on the multiple scattering.

Let us consider the first term in (\ref{2.11}) in more details:
\begin{eqnarray}
\left(\frac{\partial\hat{\rho}_{d}}{\partial
t}\right)_{col}^{(1)}=vN\frac{2\pi i}{k}
 \left[
 \hat{f}(0)\hat{\rho}_d-\hat{\rho}_d \hat{f}(0)^{+}
\right] .
\label{2.12}
\end{eqnarray}
Amplitude $\hat{f}(0)$ of forward scattering of a deuteron  in a nonpolarized target  can be expressed as:
\begin{eqnarray}
\hat{f}(0)=\emph{Sp}_{t} \hat{F(0)} \hat{\rho}_{t}.
\label{2.13}
\end{eqnarray}
In accordance with (\ref{1.2}) the amplitude reads: 
\begin{eqnarray}
\hat{f}(0)=d+d_{1}(\hat{\vec{S}}\vec{n})^{2},
\label{2.14}
\end{eqnarray}
where $\vec{n}=\vec{k}/k$, $\vec{k}$ is the deuteron momentum.
As a result the collision term governing the time evolution of the density matrix is given by:
\begin{eqnarray}
\left(\frac{\partial\hat{\rho}_{d}}{\partial t}\right)_{col}^{(1)} =
 -\frac
i\hbar\left(\hat{V} {\hat{\rho}_d}-{\hat{\rho}_d} \hat{V}^{+}\right).
\label{2.15}
\end{eqnarray}
where $\hat{V}$ is the scattering potential.

\noindent And finally, expression   (\ref{2.9}) can be presented as follows:
\begin{eqnarray}
\frac{d\hat{\rho}_{d}}{dt}=-\frac{i}{\hbar}\left[\hat{H},\hat{\rho}_{d}\right]
 -\frac
i\hbar\left(\hat{V} {\hat{\rho}_d}-{\hat{\rho}_d} \hat{V}^{+}\right){+
 vN
\emph{Sp}_{t} \int d\Omega
F(\vec{k}^{'})\hat{\rho}(\vec{k}^{'})F^{+}(\vec{k}^{'})}.
 \label{2.9_new}
\end{eqnarray}
%
%
The last term, proportional to $\emph{Sp}_{t}$, describes multiple scattering and resulting depolarization.
Hereinafter the target thickness enabling to neglect this term is used.

Beam intensity reads as
\begin{eqnarray}
I (t)=\emph{Sp}_{d}\hat{\rho}_{d}.
\label{2.16}
\end{eqnarray}
Therefore, the rate of intensity change is determined by scattering amplitude $\hat{f}(0)$ as follows:
\begin{eqnarray}
\frac{dI}{dt}=vN\frac{2\pi
i}{k}\emph{Sp}_{d}\left[\hat{f}(0)\hat{\rho}_{d}-\hat{\rho}_{d} \hat{f}^{+}(0)\right].
\label{2.17}
\end{eqnarray}
Substituting (\ref{2.7}) and (\ref{2.14}) to (\ref{2.17}) one can get 
the rate of intensity change caused by the tensor polarization components as follows
\begin{eqnarray}
\frac{dI}{dt}=\frac{\chi}{3}\left[2+p_{ik}n_{i}n_{k}\right]I(t)+\alpha
I(t),
\label{2.18}
\end{eqnarray}
where parameters  $\chi=-\frac{4\pi
vN}{k}\texttt{Im}d_{1}=-vN(\sigma_{\pm 1}-\sigma_0)$ and
$\alpha=-\frac{4\pi vN}{k}\texttt{Im}d=-v N \sigma_0$ depend on total scattering cross sections $\sigma_{\pm 1}$
and $\sigma_0$  for quantum numbers   $m= \pm 1$
and $m=0$, respectively.

Vector polarization $\vec{p}$ of a deuteron beam reads as follows:
\begin{eqnarray}
\vec{p}=\frac{\emph{Sp}_{d}\hat{\rho}_{d} \hat{\vec{S}}}{\emph{Sp}_{d}\hat{\rho}_{d}}=\frac{\emph{Sp}_{d}\hat{\rho}_{d} \hat{\vec{S}}}{I(t)}.
\label{2.19}
\end{eqnarray}
Differential equations describing vector polarization can be obtained from (\ref{2.19}):
\begin{eqnarray}
\frac{d\vec{p}}{dt}=\frac{\emph{Sp}_{d}(d\hat{\rho}_{d}/dt) \hat{\vec{S}}}{I(t)}-
\vec{p} \,\frac{\emph{Sp}_{d} (d\hat{\rho}_{d}/dt)}{I(t)}.
\label{2.20}
\end{eqnarray}
The tensor polarization components $p_{ik}$  are defined as:
\begin{eqnarray}
p_{ik}=\frac{\emph{Sp}_{d}\hat{\rho}_{d} \hat{Q}_{ik}}{\emph{Sp}_{d}\hat{\rho}_{d}}=\frac{\emph{Sp}_{d}\hat{\rho}_{d} \hat{Q}_{ik}}{I(t)},
\label{2.21}
\end{eqnarray}
where quadrupolarization operator $\hat{Q}_{ik}$ reads:
$\hat{Q}_{ik}=\frac{3}{2}\left({\hat{S}_{i}}{\hat{S}_{k}}+{\hat{S}_{k}}{\hat{S}_{i}}-\frac{4}{3}\delta_{ik}\hat{\texttt{I}}\right)$.
%
%
Similar to the vector polarization, the following equation describes evolution of the tensor polarization
\begin{eqnarray}
\frac{dp_{ik}}{dt}=\frac{\emph{Sp}_{d}(d\hat{\rho}_{d}/dt) \hat{Q}_{ik}}{I(t)}-
p_{ik}\frac{\emph{Sp}_{d} (d\hat{\rho}_{d}/dt)}{I(t)}.
\label{2.22}
\end{eqnarray}
%

\noindent Combining equations (\ref{2.7}) and (\ref{2.1}), (\ref{2.20}) and (\ref{2.22}), along with condition  $p_{xx}+p_{yy}+p_{zz}=0$, yields a system \cite{VG_2008} describing the evolution of both vector and tensor polarization components for a deutron:

\begin{eqnarray}
\left\{
\begin{array}{l}
\frac{d\vec{p}}{dt}=
\frac{e}{mc}\left[\vec{p}\times\left\{\left(a+\frac{1}{\gamma}\right)\vec{B}
-a\frac{\gamma}{\gamma+1}\left(\vec{\beta}\cdot\vec{B}\right)\vec{\beta}
\right\}\right]+\\
+\frac{\chi}{2}(\vec{n}(\vec{n}\cdot\vec{p})+\vec{p}) + \\
 +  \frac{\eta}{3}[\vec{n}\times\vec{n}^{'}]
-\frac{2\chi}{3}\vec{p}-\frac{\chi}{3}(\vec{n}\cdot\vec{n}^{'})\vec{p}
,\\
{} \\
\frac{dp_{ik}}{dt}  =  -\left(\varepsilon_{jkr}p_{ij}\Omega_{r}+\varepsilon_{jir}p_{kj}\Omega_{r}\right) + \\
 +
\chi\left\{-\frac{1}{3}+n_{i}n_{k}+\frac{1}{3}p_{ik}-\frac{1}{2}(n_{i}^{'}n_{k}+n_{i}n_{k}^{'})
+\frac{1}{3}(\vec{n}\cdot\vec{n}^{'})\delta_{ik}\right\} + \\
 +
\frac{3\eta}{4}\left([\vec{n}\times\vec{p}]_{i}n_{k}+n_{i}[\vec{n}\times\vec{p}]_{k}\right)
-\frac{\chi}{3}(\vec{n}\cdot\vec{n}^{'})p_{ik},
\\
\end{array}
\right.
\label{2.23}
\end{eqnarray}
where $\vec{n}=\vec{k}/k$,
$\eta=-\frac{4 \pi N}{k} \texttt{Re}d_{1}$,
$n_{i}^{'}=p_{ik}n_{k}$, 
$\Omega_{r}$ are the components of  $\vec{\Omega}$
($r=1,2,3$ corresponds to $x,y,z$):
\begin{eqnarray}
\vec{\Omega} & = &
\frac{e}{mc}\left\{\left(a+\frac{1}{\gamma}\right)\vec{B}
-a\frac{\gamma}{\gamma+1}\left(\vec{\beta}\cdot\vec{B}\right)\vec{\beta}
\right\} .
\label{2.24}
\end{eqnarray}

Now let us take into account that the target is located in the ring section, where the magnetic field is absent. As a result, the system of equations (\ref{2.23}) is conveniently split into two systems: one (see (\ref{2.23_a})) describes the behavior of deuteron beam spin characteristics in that section and during the time interval, where the magnetic field $\vec{B}$ is present, but the target is absent, and the second (see (\ref{2.23_b})) describes spin characteristics inside the target, where magnetic field is not applied:
\begin{eqnarray}
\label{2.23_a}
\left\{
\begin{array}{l}
\frac{d\vec{p}}{dt}=
\left[\vec{p} \times \vec{\Omega} \right]
,\\
{} \\
\frac{dp_{ik}}{dt}  =  -\left(\varepsilon_{jkr}p_{ij}\Omega_{r}+\varepsilon_{jir}p_{kj}\Omega_{r}\right)
\\
\end{array}
\right.
\end{eqnarray}

\begin{eqnarray}
\label{2.23_b}
\left\{
\begin{array}{l}
\frac{d\vec{p}}{dt}=
\frac{\chi}{2}(\vec{n}(\vec{n}\cdot\vec{p})+\vec{p}) + 
\frac{\eta}{3}[\vec{n}\times\vec{n}^{'}]
-\frac{2\chi}{3}\vec{p}-\frac{\chi}{3}(\vec{n}\cdot\vec{n}^{'})\vec{p}
,\\
{} \\
\frac{dp_{ik}}{dt}  = 
\chi\left\{-\frac{1}{3}+n_{i}n_{k}+\frac{1}{3}p_{ik}-\frac{1}{2}(n_{i}^{'}n_{k}+n_{i}n_{k}^{'})
+\frac{1}{3}(\vec{n}\cdot\vec{n}^{'})\delta_{ik}\right\} + \\
+
\frac{3\eta}{4}\left([\vec{n}\times\vec{p}]_{i}n_{k}+n_{i}[\vec{n}\times\vec{p}]_{k}\right)
-\frac{\chi}{3}(\vec{n}\cdot\vec{n}^{'})p_{ik},
\\
\end{array}
\right.
\end{eqnarray}
where $\vec{n}=\vec{k}/k$,
$\eta=-\frac{4 \pi N}{k} \texttt{Re}d_{1}$,
$n_{i}^{'}=p_{ik}n_{k}$, 
%
%
$\chi=-\frac{4\pi
	vN}{k}\texttt{Im}d_{1}=-vN(\sigma_1-\sigma_0)$.

Suppose that at instant $t_0$
a target of thickness $L$ is inserted into the beam’s path.
At this instant, particles possessing  polarization vector $\vec{p}_0$ and polarization tensor $p_{ik}^{(0)}$ pass through the boundary of the target.
After beam entering the target spin characteristics  $\vec{p}_0$ and $p_{ik}^{(0)}$ change due to interaction with the target in accordance with formulas (\ref{2.23_b}).
If the target is thin enough to make changes of vector and tensor polarization for a particle small, equations (\ref{2.23_b}) can be solved using the perturbations theory.
Therefore, from (\ref{2.23_a}) and (\ref{2.23_b})  for spin characteristics of a particle leaving the target  $\vec{p}(t_0 + \tau)$  and  
 $p_{ik}(t_0 + \tau)$ one can write:
\begin{eqnarray}
	\label{2.25_b}
\vec{p}\,(t_0+\tau)=\vec{p}_0 + 
\frac{\chi}{2}(\vec{n}(\vec{n}\cdot\vec{p}_0)+\vec{p}_0)\tau
+\frac{\eta}{3}[\vec{n}\times\vec{n}^{'}_{0 }]\tau 
-\frac{2\chi}{3}\vec{p}_0\tau
-\frac{\chi}{3}(\vec{n}\cdot\vec{n}^{'}_{0})\vec{p}_0\tau 
,
\end{eqnarray}
\begin{eqnarray}
\label{2.26_b}
\lefteqn
p_{\,~ik}(t_0+\tau)  =
p_{ik}^{(0)}
+ 
\chi\left[-\frac{1}{3}+n_{i}n_{k}+\frac{1}{3}p_{ik}^{(0)}
-\frac{1}{2}(n_{i0}^{'}n_{k}+n_{i}n_{k0 }^{'})
+\frac{1}{3}(\vec{n}\cdot\vec{n}^{'}_{0 })\delta_{ik}\right]\tau + \nonumber \\
+  \frac{3\eta}{4}\left([\vec{n}\times\vec{p}_0]_{i}n_{k}
+n_{i}[\vec{n}\times\vec{p}_0]_{k}\right)\tau
-\frac{\chi}{3}(\vec{n}\cdot\vec{n}^{'}_{0})p_{ik}^{\,(0)}\tau ,
\end{eqnarray}
where $\vec{p}_0$ is the beam polarization  at instant $t_{0}$,
$n_{i0}^{'}=p_{ik}^{\,(0)}n_{k}$,
$p_{ik}^{\,(0)}$ are the components of polarization tensor at the same instant, $\tau$ is the time interval, which the particle spends in the target.

The further evolution of $\vec{p}$ and $p_{ik}$ is again determined by the equations (\ref{2.23}).
After one revolution period $T$, a particle enters the target again possessing spin parameters $\vec{p}\,(t_0 + \tau + T)$ and
$p_{ik}(t_0 + \tau + T)$, which have changed compared to their values at the time $(t_0 + \tau)$ due to the spin rotation in the magnetic field in Nuclotron ring.
These new values can be used as the initial conditions, when solving the equations (\ref{2.23_b}), i.e. one can use the solutions of (\ref{2.25_b}) and (\ref{2.26_b}) with the replacement of $\tau$ by $\tau + T$.
This iterative process can be continued further.
However, in this case, it is more convenient to consider evolution of polarization characteristics of a particle beam in an internal target of Nuclotron in a different way.

\section{Evolution of polarization characteristics of a particle beam in an internal target of Nuclotron}
\label{sec:evolution}

To find the explicit expressions for the quantum-mechanical evolution operators, let us suppose that  $y$-axis is directed along the direction of magnetic field $\vec{B}$, and $z$-axis is parallel to the  momentum of a particle at the instant it enters a target.
Then two parts of the Hamiltonian, which are responsible for the spin dynamics of the particle in the target ($\hat V$)  can be written as follows~\cite{VG_2008,NO_EN}:
\begin{equation}
	\hat V = -\frac{2\pi \hbar^2 N}{M \gamma} \hat f(0),
\end{equation}
and the Larmor precession in the storage ring ($\hat H$) reads as~\cite{2005Mane}:
\begin{equation}
	\hat H = -\frac{e \hbar}{M c} \left( \frac{g-2}{2} + \frac{1}{\gamma} \right) B_y \hat S_y.
\end{equation}
Here $N$ is  the number of scatterers in cm$^3$ of the target, $M$ is the mass of the incident particle, $\gamma$ is its Lorentz factor, $e$ is the particle charge  and $g$ is $g$-factor (for the deuteron $g \approx 0.86$), $\hat f(0) = d_0 + d_1 \hat S_z^2$ is the amplitude of coherent elastic forward scattering in the reference frame in which the target rests.
Note that operator $\hat V$ is non-Hermitian ($\hat V \neq \hat V^+$) due to presence of nonzero imaginary parts for parameters $d_0$ and $d_1$.

Since $\tau$ is the time interval for a particle to pass through the target once, and $T - \tau \approx T$ is the time interval, when the particle moves in the storage ring beyond the target  ($\tau \ll T$), the evolution operators after passing each of two sections at a single turn are
\begin{equation}
	\hat U_V = \,{\texttt{e}}^{-i \hat V \tau / \hbar}
\end{equation}
and
\begin{equation}
	\hat U_B = \,{\texttt{e}}^{-i \hat H T / \hbar}.
\end{equation}
Then, the evolution operator for one turn in the storage ring is the product $\hat U_1 = \hat U_B \hat U_V$, and after $n$ turns it is defined as  $\hat U^{(n)} = \hat U_1^n$.

Using the following equalities valid for particles with spin $S=1$, namely: $\hat S_z^4 = \hat S_z^2$ and $\hat S_y^3 = \hat S_y$, the evolution operators $\hat U_{B(V)}$ can be transformed as follows:
\begin{equation}
	\hat U_B = \hat{\texttt{I}} + \left(\cos(\phi) - 1 \right) \hat S_y^2 + i \sin(\phi) \hat S_y
\end{equation}
and
\begin{equation}
	\hat U_V = \,{\texttt{e}}^{i \alpha} \left( \hat{\texttt{I}} + (\,{\texttt{e}}^{i \zeta} - 1) \hat S_z^2 \right),
\end{equation}
where
$$\phi=\frac{e}{mc\hbar}\left(\frac{g-2}{2}+ \frac{1}{\gamma} \right)B_y T$$
is the spin rotation angle around the magnetic field direction per single turn in the storage ring,

\begin{equation}
	\label{eq:alpha}
	\alpha=\frac{2\pi \hbar N}{M\gamma}d_0\tau
\end{equation}
is the complex quantity responsible for spin-independent beam attenuation, 
and
\begin{equation}
	\label{eq:beta}
	\zeta=\frac{2\pi \hbar N}{M\gamma}d_1\tau
\end{equation}
is responsible  for spin dichroism and spin rotation after a single pass through the target.


%
	
	Let the initial state of the particle beam in the storage ring be described by the density matrix $\hat{\rho}_0$. 
	Then the average values $p_{ij}$ of the Cartesian components of the quadrupolarization operator~\cite{VG_2008,Landau3}
	\begin{equation}
		\hat Q_{ij} = \frac{3}{2 S (S - 1)} \left( \hat S_i \hat S_j + \hat S_j \hat S_i - \frac{2}{3} \hat{\vec{S}}^{\,2} \delta_{ij} \right)
	\end{equation}
	and the average values $p_i$ of the spin operator components $\hat S_i$ read as follows:
	\begin{equation}
		\label{eq:pij}
		p_{ij} = \frac{\mathrm{Sp}(\hat \rho^{(n)} \hat Q_{ij})}{\mathrm{Sp}(\hat \rho^{(n)})}
	\end{equation}
	and
	\begin{equation}
		p_i = \frac{\mathrm{Sp}(\hat \rho^{(n)} \hat S_i)}{\mathrm{Sp}(\hat \rho^{(n)})},
	\end{equation}
	where 
	$\hat \rho^{(n)}=\hat U^{(n)}\hat\rho_0\hat U^{(n)+}$
	is the density matrix after $n$ turns of the particle in the accelerator. 
	Since the relation
	$\hat \rho^{(n)}=\hat U_1\hat\rho^{(n-1)}\hat U_1^{+}$
	holds, the matrices $\hat \rho^{(n)}$ form an explicitly defined iterative sequence, which can be constructed using mathematical software packages.
	
	In case of insignificant change of the state for an initially nonpolarized deuteron beam due to birefringence in a target
	($|\zeta n|\ll1$) the linear approximation over $\zeta$ can be used, evaluation  $\hat U_V\approx \,{\texttt{e}}^{i\alpha }
	\hat{\texttt{I}}+i\,{\texttt{e}}^{i\alpha }\zeta \hat S_z^2$ is valid and evolution operator   $\hat U$ can be approximately expressed as follows:
	\begin{equation}
		\label{eq:Ulin}
		\hat U\approx \,{\texttt{e}}^{i\alpha n}\hat U_B^n+i\,{\texttt{e}}^{i\alpha n}\Big(\sum_{j=0}^{j=n-1}{\hat U_B^{j}\zeta \hat S_z^2\hat U_B^{-j}}\Big)\hat U_B^{n-1}.
	\end{equation}
	Then at $n \gg 1$ component  $p_{zz}$ of quadrupolarization tensor  reads as follows:
	\begin{equation}
			p_{zz}\approx-  \frac{1}{3} \, n \, \texttt{Im}\zeta.
			\label{eq:pzz}
	\end{equation}
Discarding the rapidly oscillating term in \eqref{eq:pzz}, which comprises ratio of two sines, and using  \eqref{cosy_rot2+} and \eqref{eq:beta} one can get for $p_{zz}$ the following: 
	\begin{equation}
			p_{zz}\approx\frac{\Delta\sigma}{\sigma}\frac{N \sigma z}{6}.
				\label{eq:pzzlin}
	\end{equation}
	where $\Delta\sigma=\sigma_0-\sigma_{\pm 1}$ and $\sigma=\frac{2}{3}\sigma_{\pm 1}+\frac{1}{3}\sigma_0$ (for deuterons $\Delta\sigma>0$), $N$ is  the number of atoms in cm$^3$ of the target.
	Note that for an internal target $p_{zz}$ value is four times lower than for an external target at the same path length $z$ for the particle in the target (see \eqref{eq:pzzext} and \eqref{eq:pzzlin}).

Thus, deuterons passing through the Nuclotron internal target acquire tensor polarization. 
Quadrupolarization tensor component $p_{zz}$ appears to be proportional to  length $z$ of the path, which a particle passes in the target.
%
It should be noted that in case of polarization measurements for deuterons interacting with the internal target,  either deuterons scattered in the target or the products of their interactions with the nuclei in the internal target are detected rather than polarization of the transmitted beam.
For measurement with an external target, just polarization of the transmitted beam is investigated.
Particles, which came into collisions, are scattered and leave the beam.
The average value of the path length $z$ for a particle in the target is equal to mean free path $1/N\sigma$.
Therefore, the average value of $p_{zz}$ for a single cycle of Nuclotron: 
	\begin{equation}
			\bar p_{zz}\approx\frac{\Delta\sigma}{6\sigma}.
				\label{eq:pzzav}
	\end{equation}
	Further discussion is based on the application of formulas~\eqref{eq:pij} and ~\eqref{eq:pzzav} to the description of spin dichroism phenomenon for deuterons in the internal target of the storage ring.

	\section{Dichroism effect for a deuteron beam moving in Nuclotron with internal target}
	\label{sec:deuteron}

	When conducting experiments with an internal target in the storage ring, it is necessary to consider the presence of a magnetic field, which leads to Larmor precession of the deuteron spin. 
{This phenomenon leads to averaging the physical quantities and makes change in the relationship between the diagonal components of tensor polarization of the beam. }
	Due to precession, the average value of component $p_{zz}$ becomes equal to the average value of component $p_{xx}$, and due to relation $p_{xx} + p_{yy} + p_{zz} = 0$, the average value of $p_{yy}$ appears to be twice as large in magnitude as the average value of $p_{zz}$. 
	While the signs of the average values of the components $p_{zz}$ and $p_{yy}$ are opposite.
As it was already mentioned, in experiments with an internal target the detector counts scattered particles, which is in contrast to an experiment with an extracted beam, where particles transmitted through the target are detected.

\begin{figure}[ht]
		\begin{center}
			\resizebox{80mm}{!}{\includegraphics{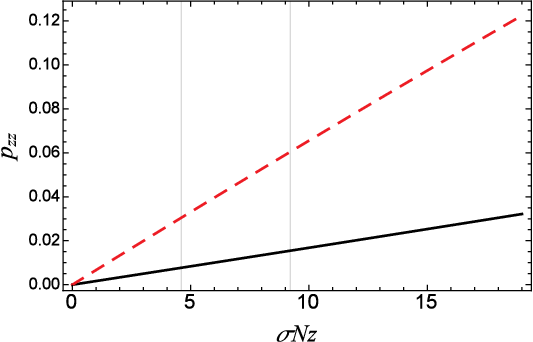}} ~~ \resizebox{80mm}{!}{\includegraphics{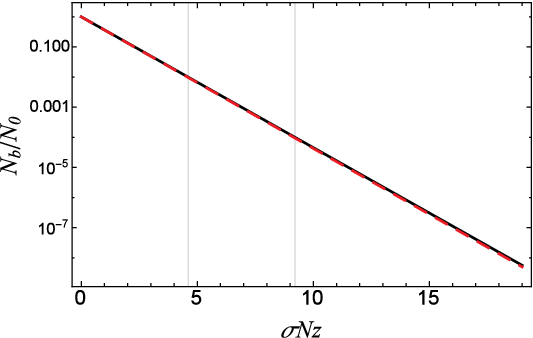}}
		\end{center}
		\caption{
Dependence of tensor polarization component $p_{zz}$  (left) and the number of deuterons in the beam (right) on the total thickness of the material layer at $\Delta\sigma / \sigma = 0.01$. Black curve corresponds to the tensor polarization of the beam in the experiments with an internal target, and gray curve is for experiments with an external target. Vertical lines on the graphs correspond to particle flux reductions by two and four orders of magnitude} \label{fig:deuteronsPzz}
	\end{figure}

	Figure~\ref{fig:deuteronsPzz} shows the dependencies of the component $p_{zz}$ and the number of particles $N_b$ in the transmitted beam 
	on the path length  $z$ for a particle in a target, 
	which is the product of the number of deuteron beam turns in the accelerator and the thickness of the target. 
	Here, the value of $\Delta\sigma / \sigma$ is taken to be $0.01$\footnote{Analysis of deuteron dichroism experiments with momentum 5 GeV$/c$~\cite{Azhgirei1,Azhgirei} shows that this value can be considerably higher ($\Delta\sigma / \sigma \approx 0.06$).}. 
	For comparison, the same graphs also show the dependence of $p_{zz}$ for the case of an external target. 
	%
	
	As an example let us consider measurements with the use of polarimeter developed by~\cite{2011Kurilkin}, which comprises polyethylene~CH$_2$ target of 10$\mu$m thickness and the deuteron beam with 270~MeV energy.  
	It is noteworthy that during a single cycle of Nuclotron operation, which duration is about several seconds, the particle beam is completely absorbed in the target
	
	For deuterons measurement of tensor polarization can be performed using the conventional approach based on measuring the angular asymmetry of deuterons elastic scattering on protons with the use of polyethylene CH$_2$ films as a target~\cite{2011Kurilkin}.
	Let us consider any four detectors (upper, lower, right, and left) arranged around the internal target of the Nuclotron at the same polar angle relative to the particle velocity direction. 
	The difference between the number of particles registered by the upper (U) and lower (D) detectors on one side and the right (R) and left (L) detectors on the other side, normalized to the total number of particles scattered into all four detectors, equals
	\begin{equation}
		\label{eq:diff}
		\frac{U + D - R - L}{U + D + R + L} = \frac{1}{2} (A_{yy} - A_{xx}) p_{zz}.
	\end{equation}
	where $A_{xx}$, $A_{yy}$ are the tensor analyzing powers for the elastic scattering reaction of deuterons on protons. 
	
	Let us consider $p_{zz}$ change in time:
	during Nuclotron operational cycle the average value $\bar{p}{zz}$ is registered. 
	This value, is approximately equal to $\bar p_{zz} \approx \frac{\Delta\sigma}{6 \sigma} \sim 2 \cdot 10^{-3}$ and includes contributions from different particles, each of them has passed through the target several (different for different particles) times and has diverse $p_{zz}$ values. 
	For example, particles, which left the beam at the very beginning of the cycle, have no tensor polarization.

	Since the angular dependence of analysing powers $A_{xx}$ and  $A_{yy}$	are well tabulated for deuteron energy 270 MeV in ~\cite{2002Satou,2011Kurilkin}, it is interesting to study analyze observables for this energy. 
	The total number of deuterons elastically scattered  by protons of polyethylene CH$_2$ within the range of polar angles from  75$^\circ$ to 135$^\circ$, and within the azimuthal size of a single detector $\sim 4^\circ$, can be evaluated as follows:
	\begin{equation}
		\label{eq:Ndet}
		N_{det} = \frac{N_0}{\sigma_{CH_2}} \int \frac{d\sigma_{el}}{d\Omega} d\Omega \approx 5 \cdot 10^{-5} N_0,
	\end{equation}
	and the difference in the particle fluxes recorded by two pairs of detectors (see \eqref{eq:diff}) reads
	\begin{equation}
		N_{diff} = \frac{N_0 \bar p_{zz}}{\sigma_{CH_2}} \int \frac{d\sigma_{el}}{d\Omega} (A_{yy} - A_{xx}) d\Omega \approx 5 \cdot 10^{-8} N_0.
	\end{equation}
	In the expressions above, $\sigma_{CH_2} \approx 800$~mb is the total scattering cross-section of a deuteron by two protons and the carbon nucleus, $\frac{d \sigma_{el}}{d \Omega}$ is the experimentally measured differential elastic scattering cross-section of deuteron by a proton taken from~\cite{2002Satou}, and $N_0$ is the {initial} number of deuterons in the beam.
	
	If the number of deuterons per a pulse is $N_0 = 10^{10}$, then $N_{diff} \approx 500$.
	At the same time, the uncertainty in determining the difference in fluxes $N_{diff}$ can be evaluated as $\Delta N_{diff} \sim \sqrt{N_{det}} \approx 700$ -- this value is of the same order as $N_{diff}$.
	Therefore, about $10^4$ {Nuclotron cycles} are required to measure tensor polarization with the  accuracy about $\sim 1\%$. 
	With a cycle duration of about several seconds, the total duration of the experiment would be around 30 hours. 
	Note that for $\Delta\sigma / \sigma \approx 0.06$~\cite{Azhgirei,Azhgirei1}, one could expect the value of $\bar p_{zz}$ to be 6 times larger ($\bar p_{zz} \approx 0.01$), and the observation time would be decreased to 1 hour.
Measurement of the tensor polarization for a deuteron beam passing through an internal carbon target at Nuclotron is of interest, since experiments with a carbon target were successfully conducted in~\cite{Azhgirei,Azhgirei1} with an extracted deuteron beam.
The use of the carbon target for such measurements
requires additional analysis of the tensor analyzing powers of reactions suitable for detecting tensor polarization.

	\section{Conclusion}

When a nonpolarized deuteron beam passes through the Nuclotron internal target, the particles acquire tensor polarization. 
According to estimations, the average value of  component $p_{zz}$ acquired by the beam particles during one accelerator cycle is approximately $\bar p_{zz} \approx 2 \cdot 10^{-3}$–$1 \cdot 10^{-2}$.
To measure the dichroism effect, it is proposed to use the existing detection system ~\cite{2011Kurilkin} based on a CH$2$ polarimeter.
Since the tensor analyzing powers of the elastic scattering process of deuterons on protons contained in the polyethylene target are rather high, the relative error in determining the average component value $\bar p{zz}$ during 30 hours of observation is about $\sim 1 \%$, assuming that  duration of  Nuclotron single cycle is several seconds.
	
The  birefringence phenomena, which is described above and reveals itself as diverse effects, namely: spin  and tensor polarization  rotation around the momentum direction, spin oscillations, vector polarization conversion to tensor that and vice versa, as well as spin dichroism, 
must be taken into account when conducting precision experiments with either nonpolarized and polarized particle beams, since they lead to changes in the components of the vector and tensor polarization of the beam and thus introduce systematic errors into the measurement results.

\section*{Acknowledgements}
\thispagestyle{empty}

The co-authors would like to acknowledge the financial support from Joint Institute for Nuclear Research (grant Belarus-JINR 234-8-2025).

\section*
	{Appendix: The Amplitude of Zero--Angle Elastic Scattering of a Deuteron by
		a Nucleus} \label{appendix}

	Let us discuss the expected magnitude of the deuteron birefringence
	effect in detail. According to (\ref{cosy_rot1}), (\ref{cosy_rot2}), (\ref{cosy_ins11}),
	the birefringence effect depends on the amplitudes of zero-angle elastic coherent
	scattering of a deuteron by a nucleus $f(m=\pm1)$ and $f(m=0)$.
	
	In order to find the amplitude $f(0)$, one should start with considering the Hamiltonian $\hat{H}$
	describing the interaction of the deuteron with the nucleus.
	
	Nonrelativistic case is considered hereinafter.
	The  Hamiltonian $H$ can be written as
	\begin{equation}
		\hat{H}=\hat{H}_D(\vec{r_p},\vec{r_n})+\hat{H}_N(\{\xi_i\})+ \hat{V}_{DN}(\vec{r_p},\vec{r_n},\{\xi_i\})\,,
		\label{eikanal3_ham1}
	\end{equation}
	where $\hat{H}_D$ is the deuteron Hamiltonian; $\hat{H}_N$ is the nuclear
	Hamiltonian; $\hat{V}_{DN}$ stands for the energy of deuteron--nucleus
	nuclear and Coulomb interaction; $r_p$ and $r_n$ are the coordinates of
	the proton and the neutron composing the deuteron, $\{\xi_i\}$ is the
	set of coordinates of the nucleons.
	
	Having introduced the deuteron
	center-of-mass coordinate $\vec{R}$ and the relative distance
	between the proton and the neutron in the deuteron
	$\vec{r}=\vec{r_p}-\vec{r_n}$, we recast (\ref{eikanal3_ham1}) as
	\begin{equation}
		\hat{H}=-\frac{\hbar^2}{2m_D}\Delta(\vec{R})+\hat{H}_D(\vec{r})+\hat{H}_N(\{\xi_i\})+\hat{V}_{DN}^N(\vec{R},\vec{r},\{\xi_i\})
		+ \hat{V}_{DN}^C(\vec{R},\vec{r},\{\xi_i\})\,, \label{eikanal3_ham2}
	\end{equation}
	where $\hat{H}_{D}(\vec{r})$ is the Hamiltonian describing the internal state of the deuteron, $m_{D}$ is the deuteron mass.
	
	In view of (\ref{eikanal3_ham2}), the deuteron--nucleus scattering
	is determined by two interactions: nuclear and Coulomb. In this
	section we shall content ourselves with finding the amplitude of
	forward elastic scattering of a deuteron with energy of hundreds
	of megaelectronvolts by a light nucleus due to nuclear interaction
	(the term $\hat{V}^{C}_{DN}$ in (\ref{eikanal3_ham2}) will be ignored).
	At lower energies, taking account of the Coulomb interaction is
	essential \cite{rins_65}.
	
	In further consideration we shall pay attention to the fact that
	for deuterons, for example,  with energy of several tens of
	MeVs, appreciably exceeding the  binding energy of
	deuterons $\varepsilon_d$, the time of nuclear deuteron-nucleus
	interaction is $\tau^N\simeq5\cdot10^{-22}$ s,  whereas the
	characteristic period of oscillation of nucleus in the deuteron is
	$\tau\simeq2\pi\hbar/\varepsilon_d\simeq2\cdot10^{-21}$\,s. So we
	can apply the impulse approximation \cite{1992VG}.  In this
	approximation we can neglect the binding energy of nucleons in the
	deuteron, i.e., neglect  $\hat{H}_D(\vec{r})$ in (\ref{eikanal3_ham2}).
	As a result,
	\begin{equation}
		\hat{H}=-\frac{\hbar^2}{2m_D}\Delta(\vec{R})+\hat{H}_N(\{\xi_i\})+\hat{V}_{DN}^N(\vec{R},\vec{r},\{\xi_i\})\,.
		\label{eikanal3_ham3}
	\end{equation}
	
	As is seen, in the impulse approximation the problem of
	determining the scattering amplitude reduces to the problem of
	scattering by a nucleus of a structureless particle having
	the same mass as the deuteron. In this case the
	coordinate $\vec{r}$ is a parameter. Therefore, the relations
	obtained for the cross section and the forward scattering
	amplitude should be averaged over the stated parameter. To
	estimate the magnitude of the effect, we shall also neglect the
	spin-dependence of internucleonic interaction. This enables using
	eikonal approximation for analyzing the magnitude of the amplitude
	for fast deuterons \cite{rins_94,quark_8}.
	
	In this approximation the amplitude of coherent zero--angle
	scattering can be written as follows:
	
	\begin{eqnarray}
		f(0)=\frac{k}{2\pi~i}\int \left( \,{\texttt{e}}^{i\chi _{D}\left( \vec{b},\vec{r}%
			\right) }-1\right) d^{2}b\left| \varphi \left( \vec{r}\right)\,
		\right| ^{2}d^{3}r, \label{eikanal3_amp}
	\end{eqnarray}
	where $k$ is the deuteron wave number, $\vec{b}$ is the impact
	parameter, $\varphi(\vec{r})$ is the wave function of the deuteron
	ground state. The phase shift due to the deuteron scattering by
	carbon is
	\begin{equation}
		\chi _{D}=-\frac{1}{\hbar v}
		\int_{-\infty }^{+\infty }V_{DN}\left( \vec{b},z^{^{\prime }},%
		\vec{r}_{\perp }\right) dz^{^{\prime }}\,,
	\end{equation}
	$\vec{r}_{\perp}$ is the component of $\vec{r}$, which is perpendicular to the
	momentum of incident deuteron, $v$ is the deuteron speed. The
	phase shift $\chi _{D}=\chi _{1}+\chi _{2}$, where $\chi_{1}$ and
	$\chi _{2}$ are the phase shifts caused by proton-nucleus and
	neutron-nucleus interactions, respectively.
	
	For the deuteron, the probability $\left| \varphi \left(
	\vec{r}\right)\right| ^{2}$ differs
	for different spin states. Thus, for
	states with magnetic quantum number $m=\pm 1$, the probability is
	$\left| \varphi_{\pm 1} \left( \vec{r}\right)\right| ^{2}$,
	whereas for $m=0$, it is $\left| \varphi_{0} \left(
	\vec{r}\right)\right| ^{2}$.
	
	Owing to the additivity of phase
	shifts, equation (\ref{eikanal3_amp}) can be rewritten as
	\begin{eqnarray}
		f\left( 0\right)=& &\frac{k}{\pi }\int \left\{
		t_{1} \left(
		\vec{b}-\frac{\vec{r}_{\perp }}{2} \right) +t_{2} \left(
		\vec{b}+\frac{\vec{r}_{\perp }}{2} \right) + 2it_{1} \left( \vec{b
		}-\frac{\vec{r}_{\perp }}{2} \right) t_{2} \left(
		\vec{b}
		+\frac{\vec{r}_{\perp }}{2} \right)
		\right\} \nonumber\\
		&\times&\left| \varphi \left( \vec{r} \right) \right| ^{2}
		d^{2}bd^{3}r\,,
		\label{eikanal3_42}
	\end{eqnarray}
	where
	\[
	t_{1(2)}=\frac{\,{\texttt{e}}^{i\chi _{1\left( 2\right) }}-1}{2i}\,.
	\]
	
	\noindent From
	(\ref{eikanal3_42}) one can get
	\begin{equation}
		f(0)=f_{1}(0)+f_{2}(0)+ \frac{2ik}{\pi}
		\int t_{1}\left( \vec{b}-%
		\frac{\vec{r}_{\perp }}{2}\right)t_{2}\left( \vec{b}+%
		\frac{\vec{r}_{\perp }}{2}\right)\left| \varphi \left(
		\vec{r}_{\perp},z\right) \right| ^{2}d^{2}bd^{2}r_{\perp}dz \,,
		\label{eikanal3_integral}
	\end{equation}
	where
	\[
	f_{1(2)}(0)=\frac{k}{\pi} \int t_{1(2)}(\vec{\xi})d^{2}\xi=
	\frac{m_D}{m_{p(n)}}~f_{p(n)}(0)
	\]
	and $f_{p(n)}(0)$ is the amplitude of the proton(neutron)--nucleus zero-angle elastic coherent scattering.
	Expression
	(\ref{eikanal3_integral}) can be recast as
	\begin{equation}
		f(0)=f_{1}(0)+f_{2}(0)+ \frac{2ik}{\pi}\int t_{1}(\vec{\xi})~
		t_{2}(\vec{\eta}) \left| \varphi
		\left(\vec{\xi}-\vec{\eta},z\right) \right|
		^{2}~d^{2}\xi~d^{2}\eta~dz\,. \label{eikanal3_27}
	\end{equation}
	Then from (\ref{eikanal3_27}), we get
	\begin{eqnarray}
		\texttt{Re}f(0)=& &\texttt{Re}f_{1}(0)+\texttt{Re}f_{2}(0) -\frac{2k}{\pi}\texttt{Im} \int
		t_1(\vec{\xi}) t_{2}(\vec{\eta})\nonumber\\
		\times& &\left| \varphi
		\left(\vec{\xi}-\vec{\eta},z\right) \right|
		^{2}~d^{2}\xi~d^{2}\eta~dz \\
		\texttt{Im}f(0)=& &\texttt{Im}f_{1}(0)+\texttt{Im}f_{2}(0)+\frac{2k}{\pi}\texttt{Re} \int t_1(\vec{\xi})
		t_{2}(\vec{\eta})\nonumber\\
		\times& &\left| \varphi \left(\vec{\xi}-\vec{\eta},z\right)
		\right| ^{2}~d^{2}\xi~d^{2}\eta~dz \,. \nonumber
		\label{eikanal3_28}
	\end{eqnarray}

	In accordance with (\ref{cosy_rot1}), (\ref{cosy_rot2}), the
	polarization state of the deuteron in the target is determined by
	the difference of the amplitudes $\texttt{Re}f(m=\pm1)$ and
	$\texttt{Re}f(m=0)$, and $\texttt{Im}f(m=\pm1)$ and
	$\texttt{Im}f(m=0)$.
	
	From (\ref{eikanal3_42}) follows that \cite{1993VG,rins_65}
	\begin{eqnarray}
		\texttt{Re}d_1=-\frac{2k}{\pi}\texttt{Im} \int t_1(\vec{\xi})
		t_{2}(\vec{\eta})\left[ \varphi_{\pm 1}^{+}
		\left(\vec{\xi}-\vec{\eta},z\right) \varphi_{\pm 1}
		\left(\vec{\xi}-\vec{\eta},z\right)\right.\nonumber\\
		\left.- \varphi_{0}^{+}
		\left(\vec{\xi}-\vec{\eta},z\right) \varphi_{0}
		\left(\vec{\xi}-\vec{\eta},z\right) \right]d^{2}\xi~d^{2}\eta~dz \\
		\texttt{Im}d_1=\frac{2k}{\pi}\texttt{Re} \int
		t_1(\vec{\xi}) t_{2}(\vec{\eta})\left[ \varphi_{\pm 1}^{+}
		\left(\vec{\xi}-\vec{\eta},z\right) \varphi_{\pm 1}
		\left(\vec{\xi}-\vec{\eta},z\right)\right.\nonumber\\
		\left.- \varphi_{0}^{+} \left(\vec{\xi}-\vec{\eta},z\right)
		\varphi_{0} \left(\vec{\xi}-\vec{\eta},z\right)
		\right]~d^{2}\xi~d^{2}\eta~dz \,,\nonumber \label{eikanal3_d1}
	\end{eqnarray}
	where $d_1$ is  the difference of spin-dependent forward scattering
	amplitudes.
	
	Note that according to (\ref{eikanal3_d1}), the spin-dependent part of the scattering
	amplitude $d_{1}$ is determined by the rescattering effects of colliding particles.
	
	When  the deuteron is scattered by a light nucleus, its
	characteristic radius  is large as compared with the radius of the
	nucleus. For this reason, to estimate the effects, we can suppose
	that in integration,
	the functions $t_1$ and $t_2$ act on $\varphi$ as a $\delta$-function. 
	Then
	\begin{eqnarray}
		\texttt{Re} d_1=& &-\frac{4\pi}{k}\texttt{Im} {f_1(0)f_{2}(0)} \int_{0}^{\infty} \left[
		\varphi_{\pm 1}^{+} \left(0,z\right) \varphi_{\pm 1}
		\left(0,z\right)\right.\nonumber\\
		& &\left.- \varphi_{0}^{+} \left(0,z\right) \varphi_{0}
		\left(0,z\right) \right]dz\,, \nonumber\\
		\texttt{Im}d_1=& &\frac{4\pi}{k}\texttt{Re}
		{f_1(0)f_{2}(0)} \int_{0}^{\infty} \left[ \varphi_{\pm 1}^{+}
		\left(0,z\right) \varphi_{\pm 1} \left(0,z\right)\right.\nonumber\\
		& &\left.- \varphi_{0}^{+} \left(0,z\right) \varphi_{0}
		\left(0,z\right) \right]dz\,. \label{eikanal3_0d1}
	\end{eqnarray}
	
	The magnitude of the birefringence effect is determined by
	the difference
	\[
	\left[ \varphi_{\pm 1}^{+} \left(0,z\right) \varphi_{\pm 1}
	\left(0,z\right)- \varphi_{0}^{+} \left(0,z\right) \varphi_{0}
	\left(0,z\right)\right]\,,
	\]
	i.e., by the difference of distributions of nucleon density in the
	deuteron for different deuteron spin orientations. The structure
	of the wave function $\varphi_{\pm 1}$ is well known
	\cite{rins_94}:
	\begin{equation}
		\varphi_m=\frac{1}{\sqrt{4 \pi}} \left\{
		\frac{u(r)}{r}+\frac{1}{\sqrt 8}\frac{W(r)}{r}\hat{S}_{12}
		\right\} \chi_m\,, \label{eikanal3_phi_m}
	\end{equation}
	where $u(r)$ is the deuteron radial wave function corresponding to
	the $S$-wave; $W(r)$ is the radial function corresponding to the
	$D$-wave; the operator $\hat{S}_{12}=6(\hat{\vec{S}}
	\vec{n}_{r})^2-2\hat{\vec{S}}^2$; $\vec{n}_{r}=\frac{\vec{r}}{r}$;
	$\hat{\vec{S}}=\frac{1}{2}(\vec{\sigma}_1+\vec{\sigma}_2)$, and
	$\vec{\sigma}_{1(2)}$ are the Pauli spin matrices describing
	proton (neutron) spin.
	
	Use of (\ref{eikanal3_phi_m}) yields
	\begin{eqnarray}
		\texttt{Re}d_1=-\frac{6}{k}~\texttt{Im} \left\{ f_{1}(0)f_{2}(0)
		\right\} G = -\frac{24}{k} \texttt{Im} \left\{ f_{p}(0)f_{n}(0)
		\right\} G\,,\
		\nonumber\\
		\texttt{Im}d_1=\frac{6}{k}\texttt{Re} \left\{ f_{1}(0)f_{2}(0)
		\right\} G= \frac{24}{k} \texttt{Re}\left\{ f_{p}(0)f_{n}(0)
		\right\} G\,, \label{eikanal3_13}
	\end{eqnarray}
	where
	\[
	G=\int_{0}^{\infty} \left( \frac{1}{\sqrt
		2}\frac{u(r)W(r)}{r^2}-\frac{1}{4} \frac{W^2(r)}{r^2} \right)
	dr\,.
	\]
	
	According to the optical theorem,
	\[
	\texttt{Im}~f_{p(n)}(0)=\frac{k_{p(n)}}{4 \pi}\sigma_{p(n)}\,,
	\]
	where $\sigma_{p(n)}$ is the total scattering
	cross section of the proton and the neutron by carbon, respectively, and 
	\[
	k_{p(n)}=\frac{m_{p(n)}}{m_{D}}k\backsimeq\frac{1}{2}k\,.
	\]
	As a result, (\ref{eikanal3_13}) can be written as
	\begin{equation}
		\texttt{Re}~d_1=-\frac{3}{\pi} \left( \texttt{Re}~f_p(0) \sigma_{n}+\texttt{Re}~f_n(0)
		\sigma_{p}
		\right) G
		\label{eikanal3a_id1}
	\end{equation}
	\begin{equation}
		\texttt{Im}~d_1=\left( \frac{24}{k}
		\texttt{Re}~f_p(0)~\texttt{Re}~f_n(0)-\frac{3k}{8
			\pi^2}\sigma_{p}\sigma_{n} \right) G\,. \label{eikanal3b_id1}
	\end{equation}
	
	In view of (\ref{eikanal3a_id1}), the analysis of the birefringence
	phenomenon in this simple approximation gives information about
	$\texttt{Re}~d_1$ and $\texttt{Im}~d_1$.
	Therefore, information about function $G$ can be obtained.

\end{document}